\documentclass[twocolumn,aps,prl,superscriptaddress,notitlepage,longbibliography,reprint]{revtex4-2}

\usepackage{graphicx}
\usepackage{epsfig}
\usepackage{float}
\usepackage{dcolumn}
\usepackage{bm}
\usepackage{amsmath}
\usepackage{float}
\usepackage{amssymb}
\usepackage{lipsum}
\usepackage{titletoc}
\usepackage{makecell}
\usepackage[colorlinks,linkcolor=blue]{hyperref}
\usepackage{array}
\usepackage{multirow}
\usepackage{color}
\usepackage{bbding}
\usepackage{stackengine}

\newcommand{\be}{\begin{equation}}
\newcommand{\ee}{\end{equation}}

\newcommand{\bea}{\begin{eqnarray}}
\newcommand{\eea}{\end{eqnarray}}

\begin{document}

\title{Fractional vortices and Ising superconductivity in multiband superconductors}

\author{Haijiao Ji} 
\author{Noah F. Q. Yuan}
\email{fyuanaa@connect.ust.hk}
\affiliation{Tsung-Dao Lee Institute, Shanghai Jiao Tong University, Shanghai 201210, China}
\affiliation{School of Physics and Astronomy, Shanghai Jiao Tong University, Shanghai 200240, China}

\begin{abstract}
Inspired by the recent experiments in monolayer iron-based superconductors, we theoretically investigate properties of a two-dimensional multiband superconductor, focusing on two aspects.
First, for vortex bound states, the spatial anisotropy and positions of electron density peaks are associated with interband couplings.
Second, even with inversion symmetry, there allows a Ising-type spin-orbit coupling, leading to the enhanced in-plane upper critical field.
\end{abstract}
\maketitle

\textit{\textcolor{blue}{Introduction.}}---
Recently, multiband superconductors with multi-component order parameters attract growing attention in the condensed matter community, with material candidates such as MgB$_2$ \cite{MgB21,MgB22,MgB23,MgB24,MgB25}, NbSe$_2$ \cite{NbSe21,NbSe22,NbSe23,NbSe24,NbSe25} and Ba$_{1-x}$K$_x$Fe$_2$As$_2$ \cite{BKFeAs1,BKFeAs2,BKFeAs3,BKFeAs4,BKFeAs5,BKFeAs6}.
It has been proposed that multiband superconductors may lead to exotic forms of superconductivity such as topological superconductivity \cite{TSC,Fetese1,Fetese2} and quantum vortices \cite{FV1,FV2}, with several experimental progress reported \cite{MgB23,MgB24,NbSe22,NbSe24,BKFeAs3,Fetese3,fraction}.

In particular, the recent experimental observation of evidence on quantum vortices carrying fractional flux quantum (so-called \textit{fractional} vortices) in iron-based superconductors by Y. Zheng, \textit{et al.} \cite{experiment} adds fuel to the study of multiband superconductivity.

Theoretically multiband superconductivity can be analyzed within the phenomenological Ginzburg-Landau framework \cite{MS,GL1,GL2,GL3,GL4,GL5}, and self-consistent calculations based on microscopic band structures \cite{self1,self2,self3,self4,self5}.
To understand the observations of fractional vortices and multiband superconductivity reported in Ref. \cite{experiment}, one first needs to analyze the microscopic band structure of iron-based superconductors.


It is generally recognized that the Fermi surfaces of iron-based superconductors mainly consist of $d$-orbital pockets of iron atoms, hybridized with orbitals from other atoms \cite{pocket1,pocket2,pocket3,pocket4,pocket5,pocket6,pocket7,pocket8}. 
In particular for KFe$_2$As$_2$ in Ref. \cite{experiment}, one may focus on hole pockets from $d_{xz}$, $d_{yz}$ and $d_{xy}$ orbitals near $\Gamma,M$ points \cite{KFeAspocket1,KFeAspocket2,KFeAspocket3,KFeAspocket4,KFeAspocket5}, and the point group is $D_{4h}$. 

In this work, we consider a two-dimensional (2D) superconductor model with $d_{xz}, d_{yz}$ and $d_{xy}$-orbital hole bands under point group $D_{4h}$. 
First, we focus on the two-band model near $\Gamma$ point formed by $d_{xz}, d_{yz}$-bands. 
After deriving the $k\cdot p$ model from symmetry analysis, we investigate the anisotropy of Fermi contours and vortex bound states. 
Then, we extend to the three-band model of $d_{xz}, d_{yz}$ and $d_{xy}$ bands, and discuss the corresponding modifications due to the additional $d_{xy}$-band. 
Finally, we self-consistently calculate the in-plane upper critical field of the multiband superconductor.

\begin{figure}
\includegraphics[width=1\columnwidth]{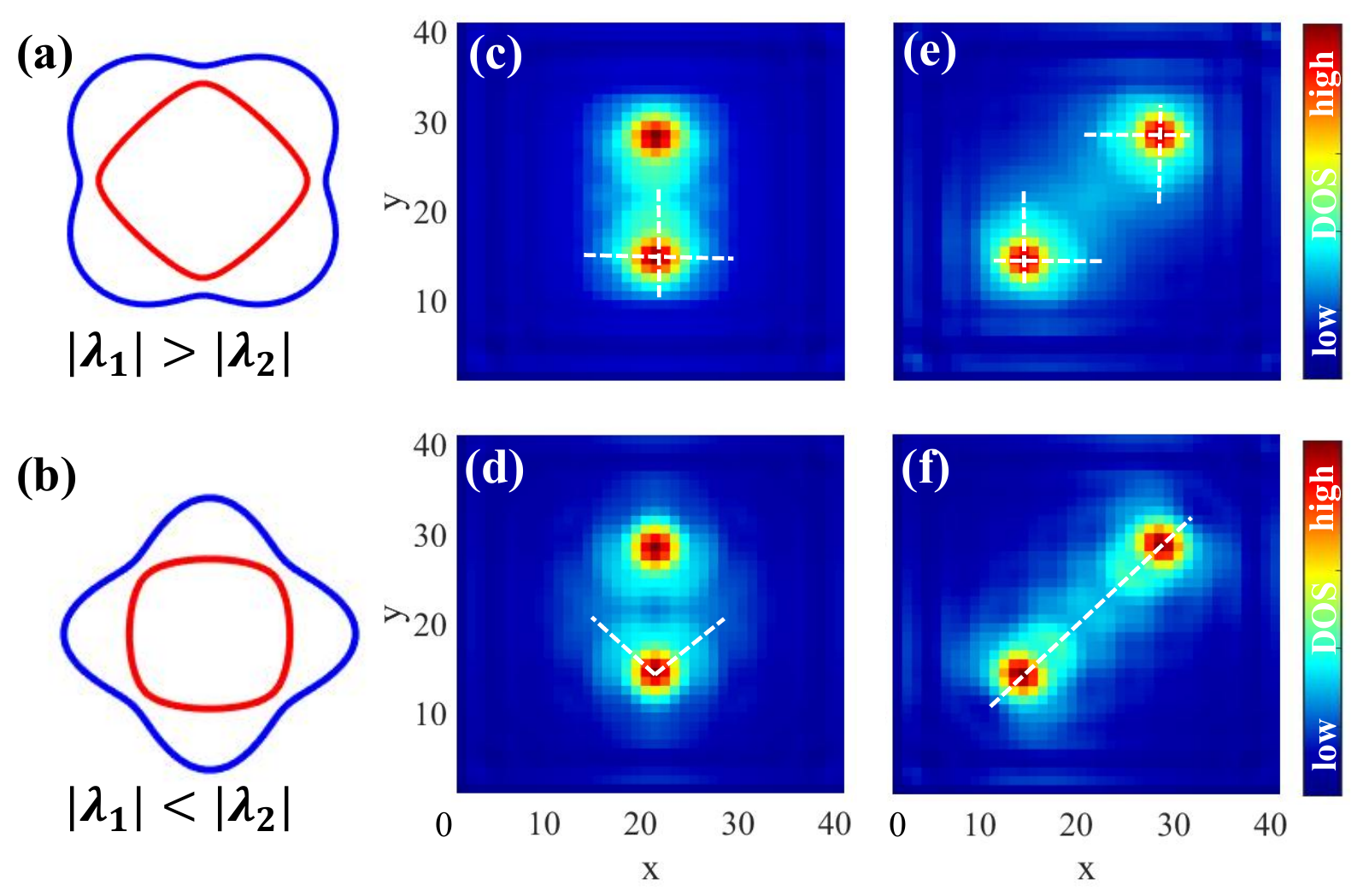}
\caption{(a, b) Fermi contours and (c-f) zero-energy density of states in two-band model Eq. (\ref{eq_HQ}). 
Dashed lines denote the directions with longer decay length. 
The two-band model Eq. (\ref{eq_HQ}) is employed with $\mu=-0.6,t=-1,\beta=0,\Delta_1=\Delta_2=0.1$.
In (c,e), $\lambda_1=0.2,\lambda_2=0.1$ and in (d,f), $\lambda_1=0.1,\lambda_2=0.2$. Fractional vortices are described by Eq. (\ref{eq_fv}) with $\xi=1$.}\label{fig1}
\end{figure}

\textit{\textcolor{blue}{Two-band model.}}---
At $\Gamma$ point, $d_{xz},d_{yz}$ orbitals furnish the 2D irreducible representation $E_g$ of point group $D_{4h}$.
In particular, under in-plane fourfold rotation $C_{4z}$, we find the eigenbasis of the orbital angular momentum $L_z$ 
\be
d_{\pm}\equiv\frac{1}{\sqrt{2}}(d_{xz}\pm id_{yz}),\quad L_z=\pm 1.
\ee

On the spinful orbital basis $\{d_{+\uparrow},d_{+\downarrow},d_{-\uparrow},d_{-\downarrow}\}$, the time-reversal symmetry $\mathcal{T}$, in-plane fourfold rotation symmetry $C_{4z}$, and vertical (horizontal) mirror symmetry $M_y$ ($M_z$) are represented as $(J_z=\sigma_z+\frac{1}{2}s_z)$
\be\label{eq_T}
\mathcal{T}=is_y\sigma_x K,\quad C_{4z}=i^{J_z},\quad
{M_y}=is_y\sigma_x,\quad M_z=is_z,
\ee
where $\bm s$ are Pauli matrices in spin space, $\bm\sigma$ in orbital space, and $K$ is the complex conjugation operator.

Since we have both orbital and spin degrees of freedom, we expect two types of couplings, namely the interband coupling (IBC) between different orbitals, and the spin-orbit coupling (SOC) between different spins.

By method of invariants,
the normal Hamiltonian up to the quadratic order of electron momentum reads
\begin{equation}
H_0(\mathbf{k}) =\varepsilon +2\lambda_1 k_x k_y\sigma_y+\lambda_2(k_x^2-k_y^2)\sigma_x +\beta s_z\sigma_z
\label{eq_HQ}
\end{equation}
where $\mathbf{k}=(k_x,k_y)$ is the electron momentum, $\varepsilon$ is the kinetic term, $\lambda_1, \lambda_2$ are spin-independent IBC parameters, and $\beta$ is the Ising SOC.  
For hole bands we have
\be
\varepsilon=-\mu-t k^2,
\ee
with chemical potential $\mu<0$ and hopping term $t<0$. 

Under time-reversal symmetry, $\bm s,\sigma_z$ are odd, while $\sigma_{x,y}$ are even. Under inversion, $\bm s,\bm\sigma$ are all even.
As a result, both IBCs and Ising SOC are even in $\mathbf{k}$, unlike those $\mathbf{k}$-odd IBCs in Benervig-Hughes-Zhang (BHZ) model\cite{SOC1} and $\mathbf{k}$-odd Rashba SOC\cite{SOC2,SOC3}.

Unlike the isotropic kinetic term $\varepsilon$ and Ising SOC $\beta$, 
IBCs $\lambda_{1,2}$ serve as the source for anisotropy in our model.
The normal Hamiltonian has the emergent symmetry
\be
(k_x,k_y)\to \frac{1}{\sqrt{2}}(k_x+k_y,k_x-k_y),\quad\lambda_1\leftrightarrow\lambda_2.
\ee
Namely, an in-plane improper rotation of angle $\pi/4$ is equivalent to the exchange of $\lambda_{1,2}$.
When $|\lambda_1|=|\lambda_2|$, the model acquires an emergent full rotation symmetry.

The anisotropy of Fermi contours in the normal Hamiltonian depends on $\lambda_1, \lambda_2$ as shown in Fig. \ref{fig1}(a) and (b).
When $|\lambda_1|>|\lambda_2|$, the outer pocket is “$\times$"-shape ($d_{xy}$-like) as in Fig. \ref{fig1}(a), and when $|\lambda_1|<|\lambda_2|$ it is “$+$"-shape ($d_{x^2-y^2}$-like) as in Fig. \ref{fig1}(b). 
In fact, with $\lambda_{\pm}^2\equiv\frac{1}{2}(\lambda_1^2\pm\lambda_2^2)$,
the bands of Eq. (\ref{eq_HQ}) read 
\begin{equation}
E_{\pm}(\mathbf{k})=\varepsilon \pm \sqrt{(\lambda_{+}^2-\lambda_{-}^2\cos 4\theta)k^4+\beta^2},
\label{eq2}
\end{equation}
in the polar coordinate $\mathbf{k}=k(\cos\theta,\sin\theta)$.


In the next session, we discuss the anisotropy effects of IBCs $\lambda_{1,2}$ in superconductivity.

\textit{\textcolor{blue}{Superconductivity.}}---
We now turn to the superconducting phase of our model, and discuss both uniform pairing and the non-uniform vortices. 

When the pairing is uniform in real space, 
we can work in the momentum space, where the electron operator $d_{\sigma s}$ carries orbital $\sigma=\pm$, spin $s=\uparrow,\downarrow$ and momentum $\mathbf{k}$, while the hole operator $d_{\sigma s}^{\dagger}$ carries orbital $-\sigma$, spin $-s$ and momentum $-\mathbf{k}$.
The Bogouliubov-de Gennes (BdG) Hamiltonian on the Nambu basis
$\{d_{+\uparrow},d_{-\uparrow},d_{+\downarrow},d_{-\downarrow},
d_{+\uparrow}^{\dagger},d_{-\uparrow}^{\dagger},d_{+\downarrow}^{\dagger},d_{-\downarrow}^{\dagger}\}^{\rm T}$ is
\begin{equation}\label{eq_BdG}
{H}_{\rm B d G}(\mathbf{k}) =
\left(\begin{array}{cc}
H_0(\mathbf{k}) & \Delta(\mathbf{k}) \\
\Delta^{\dagger}(\mathbf{k}) & -{H}^{*}_{0}(-\mathbf{k})
\end{array}\right).
\end{equation}
The pairing matrix $\Delta(\mathbf{k})$ includes spin-singlet, spin-triplet, intraband and interband channels, and the particle-hole symmetry reads
\be
\Delta(\mathbf{k})=-\Delta^{\rm T}(-\mathbf{k}).
\ee

In the following, we consider on-site pairings, which are described by an anti-symmetric pairing matrix
\be\label{eq_D}
\Delta=
\begin{pmatrix}
    0 & \Delta_{+} & \Delta_0+\Delta_z & \Delta_1 \\
    -\Delta_{+} & 0 & \Delta_2 & \Delta_0-\Delta_z \\
    -\Delta_0-\Delta_z & -\Delta_2 & 0 & \Delta_{-} \\
    -\Delta_1 & -\Delta_0+\Delta_z & -\Delta_{-} & 0 
\end{pmatrix}
\ee
with six pairing order parameters $\Delta_{1,2}$, $\Delta_{0,z}$ and $\Delta_{\pm}$, which furnish four irreducible representations of $D_{4h}$ 
\be
\Delta_{1,2}\in A_{1g},\quad
\Delta_{0}\in B_{1g},\quad
\Delta_{z}\in B_{2g},\quad
\{\Delta_{+},\Delta_{-}\}\in E_g.
\ee

In phase $B_{1g}$ or $B_{2g}$, there is only one order parameter and multiband superconductivity cannot be realized. In phase $E_g$, the two order parameters $\Delta_{\pm}$ are degenerate in the quadratic Ginzburg-Landau (GL) free energy, but have to fall into either chiral $(\Delta_{+}\Delta_{-}=0)$ or nematic $(|\Delta_{+}|=|\Delta_{-}|)$ phase when considering quartic GL terms.

We thus consider the trivial pairing phase $A_{1g}$, which is time-reversal-invariant and pairs up opposite-spin opposite-orbital states. 
It can be found that the BdG bands of Eq. (\ref{eq_BdG}) are topologically trivial, which is consistent with experimental observations in iron-based superconductors that edge states are absent \cite{experiment}.


We then move to the non-uniform pairings in the real space. 
Namely, $\Delta_{1,2}(\mathbf{r})$ can now depend on the spatial position $\mathbf{r}$.
In particular we focus on {fractional} vortices.
For simplicity, a fractional vortex for order parameter $\Delta_j$ to vanish at its core $\mathbf{r}=\mathbf{c}_j$ can be modeled by the ansatz
\be\label{eq_fv}
\Delta_j(\mathbf{r})=\Delta_j e^{i\varphi}\tanh\frac{|\mathbf{r}-\mathbf{c}_j|}{\xi},
\ee
where $\xi$ is coherent length, $\varphi$ is the polar angle of $\mathbf{r}-\mathbf{c}_j$, and $\Delta_j$ is the asymptotic pairing amplitude away from the core $|\mathbf{r}-\mathbf{c}_j|\to\infty$.
The magnetic flux trapped in such a fractional vortex is a fraction of the flux quantum and hence the name \cite{fraction,Fetese3,FV1,FV2,experiment}.

With the above ansatz, one can calculate the bound states within fractional vortices, which can be represented by the zero-energy density of states as shown in Fig. \ref{fig1}(c-f).
From the calculations, one can find that the vortex bound states (VBSs) are usually localized near the fractional vortex core with anisotropic decay length. 

It is usually found that, the VBS decay length is longer along $x$ and $y$ axes when $|\lambda_1|>|\lambda_2|$ [Fig. \ref{fig1}(c) and (e)], along diagonal directions $|x|=|y|$ when $|\lambda_1|<|\lambda_2|$ [Fig. \ref{fig1}(d) and (f)].
We may understand these in terms of effective pairing potentials on the Fermi contours $E_{\pm}=0$. It can be worked out that due to the $d$-wave-like IBCs $\lambda_{1,2}$, effective pairings on the Fermi contours are also $d$-wave-like with the form factor $-2i\lambda_1 k_x k_y+\lambda_2(k_x^2-k_y^2)$. When $|\lambda_1|>|\lambda_2|$, the effective pairings are more like $d_{xy}$-wave with $x$ and $y$ axes as weaker pairing directions, and when $|\lambda_1|<|\lambda_2|$, the effective pairings are more like $d_{x^2-y^2}$-wave with diagonal directions $|x|=|y|$ as weaker pairing directions. 
Along weaker pairing directions, VBSs will be less bounded and the decay length is longer.

The relation between anisotropy of Fermi contours and that of VBSs obtained in our ansatz-based calculations can also be found in self-consistent calculations \cite{self1}.

\textit{\textcolor{blue}{Three-band model.}}---
As mentioned previously, the additional $d_{xy}$-orbital hole band can hybridize with $d_{\pm}$-bands, and the two-band model Eq. (\ref{eq_HQ}) becomes a three-band model, which is a 6 by 6 matrix including spin.

At either $\Gamma$ or $M$ point, the point group is $D_{4h}$, and $d_{xy}$-orbital furnishes the 1D irreducible representation $B_{2g}$. In particular under $C_{4z}$ its orbital angular momentum is $L_z=2$, and under $M_y$ it changes sign. 
Since inversion symmetry is preserved in $D_{4h}$, there is no SOC for the single-band $d_{xy}$-orbital. As a result, the dispersion of $d_{xy}$-band can be described by a scalar function $\varepsilon'(\mathbf{k})$, which is $\mathbf{k}$-even kinetic term.

Near $M$ point, IBCs between $d_{xy}$-band and $d_{\pm}$-bands are negligible to the leading order, due to the large momentum/energy mismatch. We may treat $d_{xy}$-band as an additional band at $M$ point, which is decoupled from $d_{\pm}$-bands and hence does not affect the previous results in this manuscript.

Near $\Gamma$ point, due to time-reversal symmetry, the three-band Hamiltonian including SOC reads
\bea
H_0(\mathbf{k}) =
\left(\begin{array}{ccc}
h(\mathbf{k}) & 0\\
0 & h^*(-\mathbf{k})
\end{array}\right),
\eea
on the basis $\{d_{+\uparrow},d_{-\uparrow},d_{xy\uparrow},d_{-\downarrow},d_{+\downarrow},d_{xy\downarrow}\}$ with
\bea\label{eq_HT}
h =
\left(\begin{array}{ccc}
\varepsilon +\beta & \lambda & A_{+}\\
\lambda^* & \varepsilon-\beta & A_{-}\\
A_{+}^* & A_{-}^* & \varepsilon'
\end{array}\right).
\eea
To the leading order, 
we have $\varepsilon(\mathbf{k})=-\mu-t k^2$, and $\lambda(\mathbf{k})=-2i\lambda_1 k_x k_y+\lambda_2(k_x^2-k_y^2)$ as in the two-band model Eq. (\ref{eq_HQ}). The IBCs between $d_{\pm}$-bands and $d_{xy}$-band are 
\be
A_{\pm}(\mathbf{k})=a_{\pm}(k_x\pm ik_y),\quad a_{\pm}\in\mathbb{C}.
\ee
Under in-plane rotations, $d_{\pm}$-orbitals are fully isotropic, while $d_{xy}$-orbital is fourfold anisotropic. Thus, unlike fully isotropic $\varepsilon(\mathbf{k})$, we expect $\varepsilon'(\mathbf{k})$ is fourfold anisotropic
\be
\varepsilon'(\mathbf{k})=-\mu +M-t'k^2-t'' (k^4 +\eta k_x^2k_y^2).
\ee
For comparison, the fourfold anisotropy of $d_{\pm}$-bands is due to the IBC $\lambda(\mathbf{k})$ within $d_{\pm}$-bands as shown in Eq. (\ref{eq2}), and a two-band Hamiltonian Eq. (\ref{eq_HQ}) up to the quadratic terms in $\mathbf{k}$ is sufficient to describe such anisotropy.

When $\mu$ is near the gap between $\varepsilon$ and $\varepsilon'$, $|\mu|\sim |M|$, 
close to the Fermi energy, the significant bands are $d_{xy}$-band and one of $d_{\pm}$-bands,
leading to the two-band BHZ model, as derived in monolayer FeTe$_{1-x}$Se$_x$ \cite{{Fetese2}}.

\begin{figure}
\includegraphics[width=1\columnwidth]{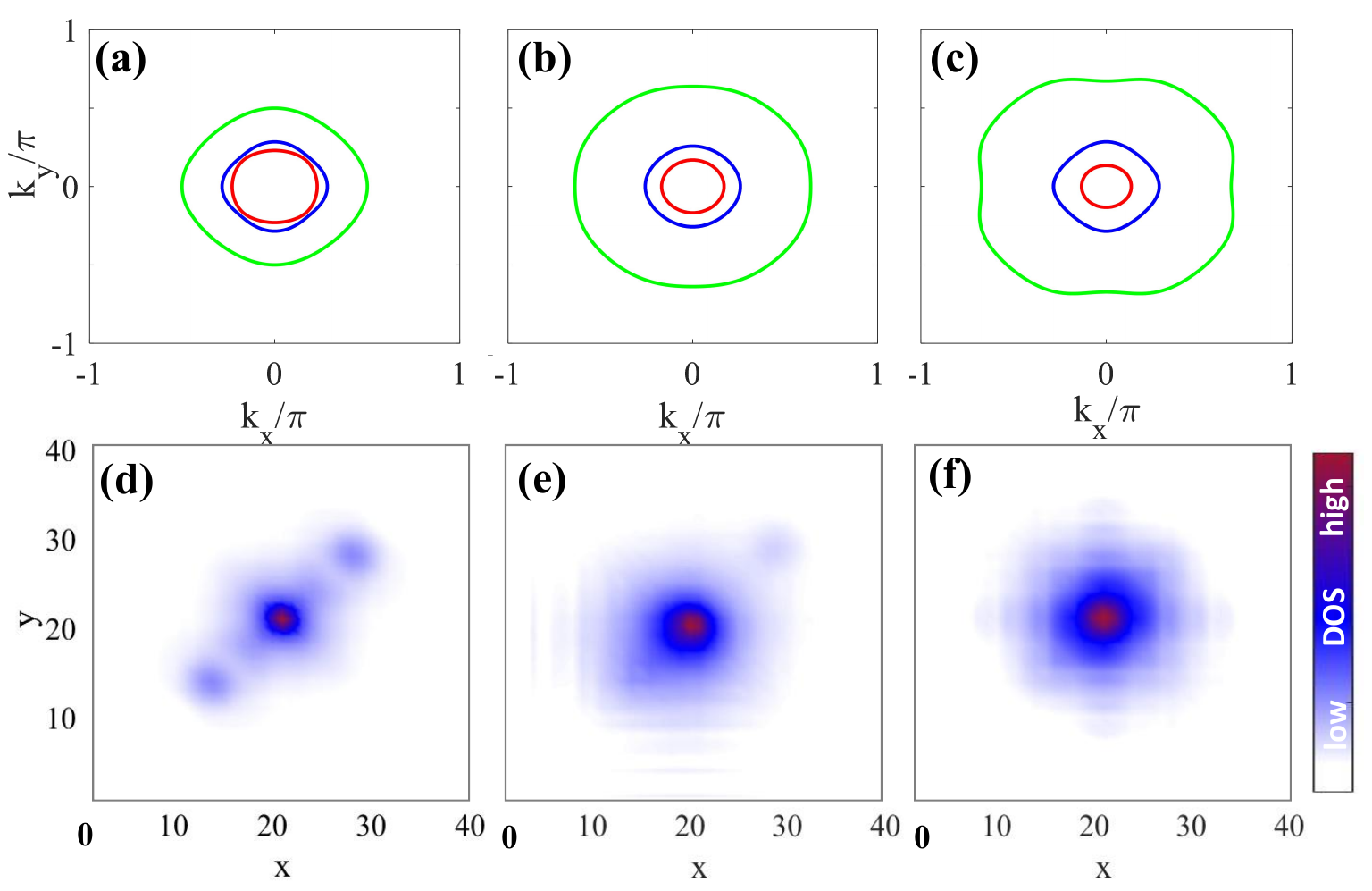}
\caption{(a-c) Fermi contours and (d-f) zero-energy density of states in three-band model Eq. (\ref{eq_HT}). Parameters are $\mu=-0.6,t=-1,\lambda_1=0.1,\lambda_2=0.2$, $t'=-0.3,t''=t'/12,\eta=-2,M=\beta=0$ and $\Delta_1=\Delta_2=0.1,\Delta_3=0.05$. In (a,d) $a_{\pm}=0$, in (b,e) $a_{+}=0$, $a_{-}=0.8$ and in (c,f) $a_{\pm}=0.8$. In (d-f), three fractional vortices of $\Delta_{j}(\mathbf{r})$ as modeled by Eq. (\ref{eq_fv}) with $j=1,2,3$ and $\xi=1$ are placed at $\mathbf{c}_1=(28,28),\mathbf{c}_2=(12,12)$, and $\mathbf{c}_3=(20,20)$, leading to 3 (d), 2 (e) and 1 (f) peaks in zero-energy density of states respectively.}\label{fig2}
\end{figure}

On the contrary, deep in the hole band, $|\mu|\gg|M|$, the three-band model Eq. (\ref{eq_HT}) has to be employed.
In this case, one may investigate the role of IBCs in the Fermi contour anisotropy. 
It can be found that, when $\lambda_1+\lambda_2=0$ and $\eta=0$, all three Fermi contours are circular and fully isotropic in the three-band model Eq. (\ref{eq_HT}), which is a special case of the isotropic condition $|\lambda_1|=|\lambda_2|$ for the two-band model Eq. (\ref{eq_HQ}). In general the three Fermi contours are fourfold anisotropic when $\lambda_1+\lambda_2\neq 0$ and $\eta\neq 0$, as shown in Fig. \ref{fig2}(a-c), where we plot Fermi contours with fixed $\lambda_{1,2}$ but different $a_{\pm}$.


In the superconducting phase $A_{1g}$, besides $\Delta_{1,2}$ for $d_{\pm}$-bands, a trivial $s$-wave spin-singlet pairing order parameter $\Delta_3$ for $d_{xy}$-band is also included.
The pairing Hamiltonian is described by on-site pairings 
\be\label{eq_D}
H_{A_{1g}}=\sum_{\mathbf{r}}\Delta_{1}d_{+\uparrow}d_{-\downarrow}+\Delta_{2}d_{-\uparrow}d_{+\downarrow}+\Delta_{3}d_{xy\uparrow}d_{xy\downarrow}+h.c.
\ee
where the sum is over all lattice sites $\mathbf{r}$.

Under an out-of-plane magnetic field, fractional vortices of $\Delta_{j}(\mathbf{r})$ as modeled by Eq. (\ref{eq_fv}) with $j=1,2,3$ and the corresponding VBSs can be induced. 
In Fig. \ref{fig2} (d-f) we calculate and plot the zero-energy density of states.
When all IBCs are weak, one may expect three peaks localized at the cores of three fractional vortices as shown in Fig. \ref{fig2}(d). 
As the IBCs increase, VBS peaks at nearest fractional vortices may merge together, resulting in two [Fig. \ref{fig2}(e)] or even one peak [Fig. \ref{fig2}(f)]. 
Thus, when one vortex is split into three fractional vortices, the observed number of VBS peaks can be one, two or three, depending on the details of IBCs. 
Our theoretical calculations are consistent with the experiments \cite{experiment}, where VBSs with 1, 2 and 3 peaks have all been observed.


The fractional vortices are induced and stabilized by out-of-plane magnetic fields. In the following we will consider effects of in-plane magnetic fields.


\textit{\textcolor{blue}{In-plane upper critical field.}}---
When an in-plane magnetic field $\bm B$ is applied, the orbital effect can be neglected and the Zeeman effect dominates. 
In the normal state, the three-band Hamiltonian Eq. (\ref{eq_HT}) is modified to contain the Zeeman coupling term
\begin{equation}
H_0(\mathbf{k}) \to H_0(\mathbf{k}) +\frac{1}{2}\mu_{\rm B} \hat{g}\bm B\cdot\bm s,
\end{equation}
where $\mu_{\rm B}$ is the Bohr magneton, $\hat{g}={\rm diag}(g_{+},g_{-},g_3)$ is a matrix in orbital space and $g_{\pm},g_3$ are the Land\'e $g$-factors for $d_{\pm}$-orbitals and $d_{xy}$-orbital respectively. Due to time-reversal symmetry, $g_{+}=g_{-}\equiv g$.

In the superconducting state, we need to focus on the Fermi contours where pairing occurs.
On the Fermi contour of $d_{xy}$-band, due to inversion symmetry, there is no SOC in the orbital singlet $d_{xy}$-band. As a result, the in-plane upper critical field is $2B_{c2}^{\rm BCS}/g_3$, where $B_{c2}^{\rm BCS}$ is the critical field of a Bardeen-Cooper-Shriffer (BCS) superconductor purely due to Zeeman effect.

On the Fermi contour of $d_{\pm}$-bands, since $d_{\pm}$-orbitals form a doublet, there exists Ising SOC as shown in Eqs. (\ref{eq_HQ}) and (\ref{eq_HT}).
Similar to the effective pairing potentials, the effective Zeeman field on $d_{\pm}$-bands 
can be obtained 
\be\label{eq_Beff}
\bm B_{\rm eff}=\frac{\lambda_{\rm F}}{\sqrt{\lambda_{\rm F}^2+\beta_{\rm F}^2}}\bm B,
\ee
where $\lambda_{\rm F}$ and $\beta_{\rm F}$ are angular averages of IBC and SOC respectively along three Fermi contours (FCs)
\be\label{eq_lb}
\lambda_{\rm F}=\int_{\rm FC}\frac{d\theta}{2\pi}|\lambda(\mathbf{k})|,\quad
\beta_{\rm F}=\int_{\rm FC}\frac{d\theta}{2\pi}|\beta(\mathbf{k})|,
\ee
with $\theta$ as the polar angle of $\mathbf{k}$.
According to our previous assumption, we have
$|\mu|\gg \lambda_{\rm F},\beta_{\rm F},\Delta_{1,2}.$

\begin{figure}
\includegraphics[width=1\columnwidth]{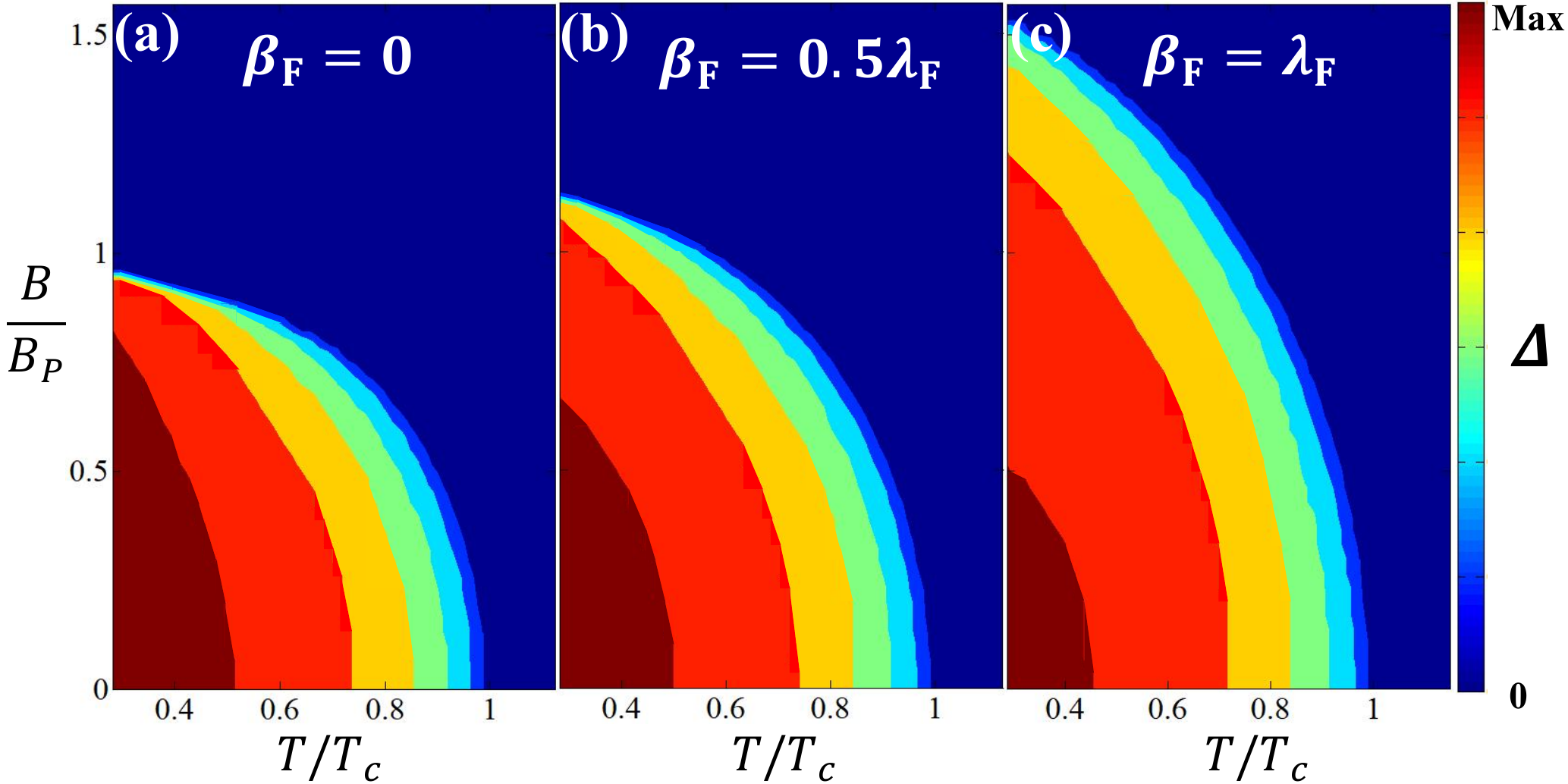}
\caption{Phase diagrams with different Ising SOCs. The two-band model Eq. (\ref{eq_HQ}) is employed with $\mu=-0.6,t=-1,\lambda_1=0.1,\lambda_2=0.2$, and (a) $\beta=0$ (b) $\beta=0.1$ and (c) $\beta=0.2$.}\label{fig3}
\end{figure}

From Eq. (\ref{eq_Beff}), the in-plane upper critical field $B_{c2}$ is
\be\label{eq_Bc2}
\frac{B_{c2}}{B_{c2}^{\rm BCS}}=\frac{2}{g}
\sqrt{1+\left(\frac{\beta_{\rm F}}{\lambda_{\rm F}}\right)^2}.
\ee
In Fig. \ref{fig3} we work out and plot the self-consistent pairing gap $\Delta=\Delta_1=\Delta_2$ as a function of temperature $T$ and field $B$. In the phase diagram, the phase transition line denotes $B_{c2}$.
Without Ising SOC, $B_{c2}$ becomes the BCS critical field $2B_{c2}^{\rm BCS}/g$, as shown in Fig. \ref{fig3}(a), which increases to the well-known Pauli limit $B_P$ at zero temperature.
As shown in Fig. \ref{fig3}(b) and (c), when Ising SOC is nonzero, $B_{c2}$ can bypass the Pauli limit even at nonzero temperatures, and the field enhancement increases with the ratio ${\beta_{\rm F}}/\lambda_{\rm F}$ as described in Eq. (\ref{eq_Bc2}).

Physically, the inversion symmetry is preserved, and the Ising SOC couples spin and orbitals near $\Gamma$ point. For each orbital, Ising SOC behaves like a Zeeman field along out-of-plane direction, which enhances $B_{c2}$. For opposite orbitals, Ising SOCs are opposite in sign but the same in magnitude, to preserve time-reversal symmetry.
However, IBCs $\lambda_{1,2}$ couple opposite orbitals and tend to weaken the spin pinning of Ising SOC for each orbital. As a result, we arrive at the enhancement formula Eq. (\ref{eq_Bc2}) for the in-plane upper critical field. This is one example of Ising superconductivity \cite{JMLu,XXi,JF,HL}.

In our calculations, we assume the Cooper pair momentum is always zero.
At low temperatures, finite-momentum pairing states such as Fulde-Ferrell-Larkin-Ovchinnikov phase may arise beyond Pauli limit \cite{FF,LO}, which could further enhance the in-plane upper critical fields.
In particular, in the Fulde-Ferrell phase the inversion symmetry can be spontaneously broken, leading to the potential supercurrent diode effect \cite{Yuan,Akito,James}.

\textit{\textcolor{blue}{Conclusion.}}---
In this work, we study the two- and three-band models arising from iron-based superconductors. 
We analyze the properties of Fermi contours and vortex bound states, which are related to the couplings between different bands.
Our results may help to explain the experimental observation of vortex bound states in monolayer iron-based superconductors.
We also calculated the in-plane upper critical field self-consistently, and predict Ising superconductivity might be observed in monolayer iron-based superconductors such as Ba$_{1-x}$K$_x$Fe$_2$As$_2$, LiFeAs and FeTe$_{1-x}$Se$_x$.


\textit{\textcolor{blue}{Acknowledgements.}}---
We thank Egor Babaev for important discussions.
We thank Hong Ding, Baiqing Lv and Quanxin Hu for helpful discussions.
This work is supported by the National Natural Science Foundation of China (Grant. No. 12174021).

\end{document}